\documentclass[manuscript]{aastex631}

\begin{document}

\title{Excited-State OH Masers in the Water Fountain
Source IRAS~18460$-$0151}

\correspondingauthor{Yong Zhang}
\email{zhangyong5@mail.sysu.edu.cn}

\author[0000-0002-2762-6519]{Xu-Jia Ouyang}
\affiliation{School of Physics and Astronomy, Sun Yat-sen University, 2 Daxue Road, Tangjia, Zhuhai, Guangdong Province,  China}

\author[0000-0002-1086-7922]{Yong Zhang}
\affiliation{School of Physics and Astronomy, Sun Yat-sen University, 2 Daxue Road, Tangjia, Zhuhai, Guangdong Province,  China}
\affiliation{CSST Science Center for the Guangdong-Hongkong-Macau Greater Bay Area, Sun Yat-Sen University, Guangdong Province, China}

\author[0000-0003-3520-6191]{Juan Li}
\affiliation{Department of Radio Science and Technology, Shanghai Astronomical observatory, 80 Nandan RD, Shanghai 200030, China}
\affiliation{Key Laboratory of Radio Astronomy, Chinese Academy of Sciences, China}

\author[0000-0003-3324-9462]{Jun-ichi Nakashima}
\affiliation{School of Physics and Astronomy, Sun Yat-sen University, 2 Daxue Road, Tangjia, Zhuhai, Guangdong Province,  China}
\affiliation{CSST Science Center for the Guangdong-Hongkong-Macau Greater Bay Area, Sun Yat-Sen University, Guangdong Province, China}
                
\author[0000-0002-5435-925X]{Xi Chen}
\affiliation{Center for Astrophysics, Guangzhou University, Guangzhou 510006, China}
\affiliation{Shanghai Astronomical Observatory, Chinese Academy of Sciences, 80 Nandan Road, Shanghai 200030}

\author[0000-0003-0196-4701]{Hai-Hua Qiao}
\affiliation{National Time Service Center, Chinese Academy of Sciences, Xi'An, Shaanxi 710600, People's Republic of China}
\affiliation{Key Laboratory of Time Reference and Applications, Chinese Academy of Sciences, China}



\begin{abstract}

Water fountain objects are generally defined as ``evolved stars with low to intermediate initial mass accompanied by high-velocity molecular jets detectable in the 22.235\,GHz H$_2$O maser line''. They are the key objects of
understanding the morphological transitions of circumstellar envelopes
during the post asymptotic giant branch phase.
Masers are useful tools to trace the kinematic
environments of the circumstellar envelopes.
In this letter we report the discovery of exceptionally uncommon excited-state hydroxyl (ex-OH) masers at 4660 and 6031\,MHz toward the water fountain source IRAS~18460$-$0151. 
These are the brightest ex-OH masers discovered in late-type objects to date.
To the best of our knowledge, prior to the current work, no evolved stellar object has been observed in the 4660\,MHz ex-OH maser line. 
The ground-state hydroxyl (g-OH) masers at 1612 and 1665\,MHz are also observed.
The velocity components of the 4660\,MHz
ex-OH maser line and the much weaker 1665\,MHz g-OH maser line all can be seen in
the 1612\,MHz g-OH maser line profile. 
{ The blue-shifted components of the three masers are more intense than the red-shifted ones, in contrast to the ex-OH maser line at 6031\,MHz.}
The relevance of the  behaviors of the ex-OH masers to the circumstellar
environments is unclear.

\end{abstract}

\keywords{Astrophysical masers (103); Hydroxyl masers (771); Water masers (1790); Asymptotic giant branch stars (2100); Circumstellar masers (240)}

\section{Introduction} \label{sec:intro}

During their late evolutionary stage, low and intermediate mass stars (0.8--8\,${\rm M}_\sun$) will evolve to an asymptotic giant branch (AGB) phase, at which phase the star  becomes incredibly unstable, with material flowing out and accumulating around the star to form a circumstellar envelope (CSE).
The spherically symmetric CSE may transition into a bipolar or multipolar 
nebula
after a short proto-planetary nebula (proto-PNe) period with a timescale
of  $\sim$10$^2$--10$^4$\,yr \citep[e.g.,][]{1993ARA&A..31...63K,2000ApJ...528..861U,2001MNRAS.322..321G,2016MNRAS.461.3259G}.
Despite years of intensive studies \citep[e.g.,][]{1998AJ....116.1357S},
it remains unclear how the PNe are shaped. 
Many processes have been proposed to generate the bipolar morphology.
A widely accepted scenario is that a common envelope developed by the binary interaction produces a high pole-to-equator density contrast which
 interacts with the subsequent  fast wind to form the bipolar shaped nebulae 
 \citep{1988ApJ...329..764L}.
Apart from that, 
  magnetic fields might play some role on affecting the mass-loss rate, sculpting the stellar wind, and aligning the outflows 
\citep[e.g.,][]{2002ApJ...576..976F,2004ApJ...614..737F,2006MNRAS.370.2004N,2014MNRAS.439.2014T,2020ApJ...889...13B}.
However, it is  unclear
how magnetic fields are generated and change
during the evolution from the AGB to proto-PN to PN stages.

One of the primary approaches of measuring the magnetic field in evolved stars is the polarization measurement of masers \cite[e.g.,][]{2014IAUS..302..389V}.  
Ground-state OH masers (g-OH; 1.6–1.7\,GHz), water masers (22.2\,GHz), and SiO masers (43.1\,GHz) have been commonly discovered in oxygen-rich AGB CSEs,
which have been 
 considered as crucial tracers of nebular conditions.
Comparing the spatial distribution of the SiO maser spots and 
the orientation of the magnetic field in an OH/IR star, \cite{2012A&A...538A.136A}
concluded that the stellar outflow was shaped and defined by a dipole magnetic field.

``Water fountain" stars (WFs) are evolved stars equipped with high-velocity water masers \citep[see][for reviews]{2007IAUS..242..279I,2012IAUS..287..217D}. 
Their water maser spectra have a typical velocity spread of $\ge$100\,km\,s$^{-1}$ and a maximum velocity spread of $\simeq$500\,km\,s$^{-1}$ \citep{2011ApJ...739L..14G}. 
Thus far, 15 WFs have been found by interferometry measurements \citep{2012IAUS..287..217D,2017MNRAS.468.2081G}. 
The spatiokinematics of the water masers of the WFs exhibit a broad diversity \citep{2020PASJ...72...58I,2023ApJ...948...17U}.
Given the short dynamical timescales of maser jets \citep[5-100 yr;][]{2007IAUS..242..279I,2020ApJ...890L..14T,2023PASJ...75.1183I} and  high optical obscuration \citep{2008ApJ...689..430S}, WFs are likely to be in
a transient phase between the AGB and proto-PN periods \citep{2022NatAs...6..275K}. 
Consequently, WFs could be one of the first manifestation of 
collimated mass-loss in evolved stars, and thus are
important objects for determining the timing and mechanisms of the disruption of spherical symmetry of the CSE.

The strength and direction of the in-situ magnetic fields 
in WFs could be measured in terms of the OH masers. 
As higher column densities are required for 
the inversion of the ex-OH line \citep{2002MNRAS.331..521C,2007IAUS..242..336W},
 g- and ex-OH masers are able to trace the
 magentic fields at different circumstellar locations.
However, there have been very few observations of ex-OH masers in CSEs 
\citep{1996A&ARv...7...97H}.
Unambiguous detections include the 4765\,MHz maser line in
CRL~618 \citep{2019ApJ...878...90S} and the 6035\,MHz maser line
in Vy\,2-2 and K\,3-35 \citep{2010A&A...520A..45D}.
Less confident detections of the 4751\,MHz maser line in AU~Gem and
the 6031 and 6035\,MHz lines in NML~Cyg have been reported \citep{1972ApJ...177...59Z,1981ApJ...250L..77C}, 
which were not confirmed by subsequent observations probably due to
either time variation in intensity or false signal \citep[][and references therein]{2007ApJ...666L.101S}. 
In this letter, we report the first detection of ex-OH masers toward 
a oxygen-rich object IRAS~18460$-$0151 (I18460 hereafter). 

I18460, firstly identified as  a WF source by \cite{2007ApJ...664.1130D}, is a poorly studied target. 
It does not manifest 
radio continuum emission at 6\,cm 
\citep{1985A&A...144..514H}.
Its water maser spectrum shows
a velocity spread of $\sim$290\,km\,s$^{-1}$ with a central velocity of $\sim$100\,km\,s$^{-1}$, while the CO ($J=1-0$ and $J=2-1$) lines peak at $\sim$125\,km\,s$^{-1}$ \citep{2013A&A...560A..82R}.
The g-OH maser at 1612\,MHz exhibits two peaks \citep{1997A&AS..124..509S} with velocities at 111 and 138.3\,km\,s$^{-1}$. 
The SiO $J=1-0$ and $J=2-1$ transitions were not detected \citep{2007ApJ...664.1130D}. 
Recent observations of CO $J=2-1$ and C$^{18}$O $J=2-1$ by the Atacama Large Millimeter/submillimeter Array (ALMA) indicate
a lifetime of 150 years for the envelope \citep{2022NatAs...6..275K}.

\section{Observations and Data Reduction} \label{sec:obse}

The observations were performed at the Tianma  65-m Radio Telescope (TMRT), 
operated by Shanghai Astronomical Observatory, in the `position switching' mode
with an OFF position displaced by 0.5$\arcdeg$ in right ascension.
The pointing position was $\rm R.A. = 18^h48^m43^s.02$, $\rm decl. = -01\arcdeg48\arcmin30\arcsec.2$ (J2000).
The cryogenically cooled L/C/K band receivers were utilized for the
observations of OH and water masers.
The Digital Backend System was used to record signals in the left-hand circular polarization (LCP) and the right-hand circular polarization (RCP), simultaneously \citep{2016ApJ...824..136L}.
The signal of noise diodes was injected for the flux calibration.
The calibration uncertainty is estimated to be within 30\%.

The L-band observations were performed  on December 16th and 18th 2023 with an integrated time of 20 minutes each. For the backend, Mode 24, which provides a bandwidth of 23.44\,MHz in  65,536 channels, was used to record the 
OH masers at 1612, 1665, 1667, and 1720\,MHz at four individual spectral windows. The velocity resolution is $\sim$0.08\,$\rm km~s^{-1}$ at 1.6\,GHz.
The aperture efficiency of the telescope is $\sim$55\%, corresponding to 
a sensitivity of $\sim$1.5\,$\rm Jy\,K^{-1}$. The system temperature is about 20--30\,K. The half-power beamwidth (HPBW) is $\sim$10$\arcmin$ at 1.6\,GHz.

We carried out the C-band observations on December 10th, 16th, and 18th 2023 with an integrated time of 32, 20, and 64 minutes, respectively. Mode 24
of the backend was used, allowing to simultaneously cover the seven ex-OH transition lines at 4660, 4751, 4766, 6017, 6031, 6035, and 6049\,MHz. The velocity resolution is 
$\sim$0.02\,$\rm km~s^{-1}$ at 6.0\,GHz.
In order to improve the signal-to-noise ratio, 
the spectra covering the ex-OH maser at 6031\,MHz were smoothed
to reduce the resolution to 0.07\,$\rm km~s^{-1}$.
The aperture efficiency of the telescope, the sensitivity, and the
system temperature are the same as those of L-band observations.
The HPBW is $\sim$2$\arcmin$.6 at 6.0\,GHz.

The K-band observations were conducted on 2023 December 18 with an
integration time of 16 minutes. 
The spectrum was recorded using Mode 20, which provides a bandwidth of  23.44\,MHz in 4096 channels, resulting in a velocity resolution of
0.08\,$\rm km~s^{-1}$ at 22\,GHz.
The  aperture efficiency was about 50\%, corresponding to a sensitivity of $\sim$1.7\,$\rm Jy\,K^{-1}$.
The HPBW is $\sim$0$\arcmin$.7 at 22.0\,GHz.
 The water maser spectrum
 was smoothed to a resolution
 of 0.3\,$\rm km~s^{-1}$.

The spectral data was processed using the GILDAS/CLASS  \footnote{See \url{http://www.iram.fr/IRAMFR/GILDAS} and \citet{2005sf2a.conf..721P}.} package. 
A detection  was considered to be real only if its flux density 
was stronger than 3$\sigma$ root-mean-square (rms) noise level, the signal spanned more than
three adjacent channels, and both LCP and RCP exhibited the signal.
The spectra taken on different dates
are co-added to measure the maser lines. 
Each line exhibits one or more separate
groups consisting of blended components.
The measurement results are presented in Table~\ref{tab:pro}, where each row gives 
those of each group, 
$\int S_vdv$ is the integrated flux density,
V$_{\rm p}$ and S$_{\rm p}$ are the peak velocity
and flux density, and the column
titled with `Range'
gives the minima and maxima velocities
of the group.
Because most of the groups exhibit
multiple blended components, we 
opted not to 
perform Gaussian fittings.
The recent CO observations of I18460 suggest that
the molecular gas is confined with a radius of
 0$\arcsec$.96 \citep{2022NatAs...6..275K},
 which allows us to estimate the lower limits of the brightness temperatures
(T$_{\rm b}$), as listed in the last column of Table~\ref{tab:pro}.

\section{Results and Discussion} \label{sec:resu}

\begin{figure}[ht!]
\plotone{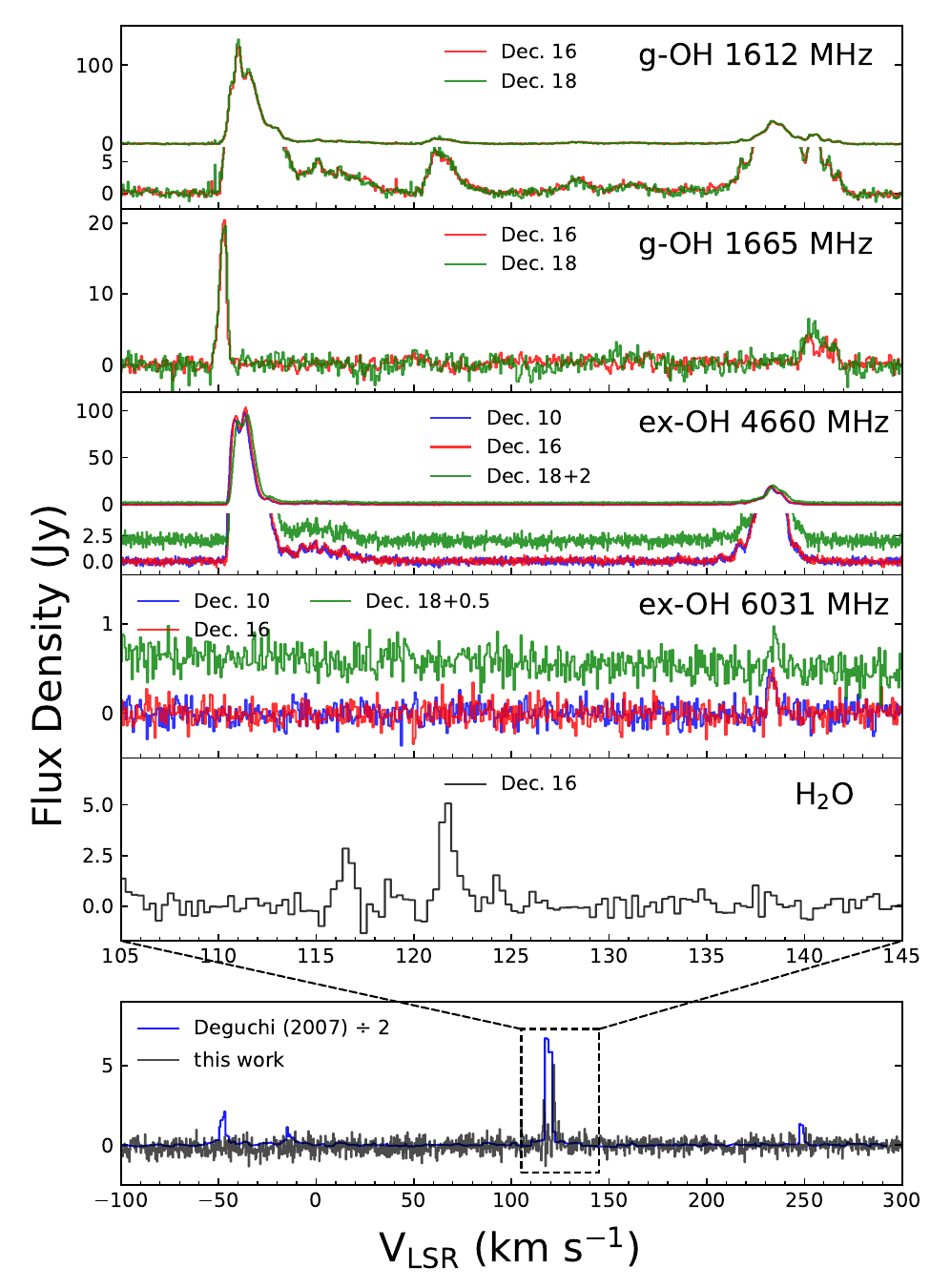}
\caption{The maser spectra of I18460 after subtracting the baseline. 
The flux density is shown
for the Stokes I (LCP+RCP).
In each panel, the colorful curves represent the spectra taken on different dates. 
Zoom-in views are provided below the
spectra of the 1612 and 4660 lines.
The bottom panel compares the water maser line observed in this work and that by \citet{2007ApJ...664.1130D}
in a broader velocity range.
\label{fig:oh}}
\end{figure}

Figure~\ref{fig:oh} shows the Stokes I (LCP$+$RCP)
spectra of the detected masers.
The spectra taken on
different dates appear largely similar. We do not see any time variation in the
flux and frequency of the detected features.
The g-OH maser lines at 1612 and 1665\,MHz are clearly detected 
and the intensity ratio of the 1612 and 1665\,MHz lines is consistent with the properties of OH masers in typical evolved stars \citep{1996A&ARv...7...97H}. 
But that at 1667\,MHz is invisible at a noise level of $\sim300$\,mJy.
Both 1612 and 1665\,MHz masers exhibit two well-separated velocity
components at velocities of about 110 and 140\,$\rm km~s^{-1}$, while the former shows
several sub-features and lower-velocity components. 
A close inspection of Figure~\ref{fig:oh} shows
that the 1665\,MHz maser is slightly more extended 
than the 1612\,MHz maser in the velocity range (see also Table~\ref{tab:pro}). This behavior is opposite
to that of most of the AGB envelopes, where the 
1612\,MHz maser origins from the outermost regions
of the accelerated envelope. A possible explanation 
is that the inner regions may have been accelerated 
through the interaction with the WF jets situated within
the OH shell.
High spatial resolution observations and an
elaborated  modelling are required to
clarify this.

The ex-OH masers at 4660 and 6031\,MHz are 
unambiguously detected in the spectra of all dates.
No other late-type stars exhibit so intense ex-OH masers.
The  4660\,MHz line arises from the
$^2\prod_{1/2}$, $J=2/1$, $F=0$--1 transition. 
The blue-shifted and red-shifted components of the 4660\,MHz ex-OH maser peak at 111.4 and 138.3\,$\rm km~s^{-1}$, respectively,
with a central velocity of about 124.9\,$\rm km~s^{-1}$.
Both the 4660\,MHz and 1612\,MHz masers
exhibit  sub-features at the red wing of the 
111.4\,km\,s$^{-1}$ group and at the blue wing of the 138.3\,km\,s$^{-1}$ group. 
We note that the interferometric 
observations of the molecular cloud
Sgr B shows
that the 4660\,MHz and 1612\,MHz maser spots 
are spatially coincident \citep{1987MNRAS.225..469G} . 
However,  weak sub-features
at 120--135\,km\,s$^{-1}$ and 141--143\,km\,s$^{-1}$ of
the 1612\,MHz  maser are invisible in the 4660\,MHz maser.
The blue-shifted group of the 4660\,MHz
maser has a flux density of 103.4\,Jy, which is much stronger than those detected in  high-mass star-forming regions \citep[HMSFR,][]{2022ApJ...928..129Q}.
The weak 6031\,MHz maser, arising from the
$^2\prod_{3/2}$, $J=5/2$, $F=2$--2 transition, shows only one red-shifted component peaking at $138.3$\,$\rm km~s^{-1}$ with a sharp profile. 
Under a local thermodynamic equilibrium condition,
the intensity ratio between the 4660\,MHz and the 6031\,MHz lines should
be $\sim1.5$. However, the measured values are $38.4$ and $>2000$
for the red-shifted and blue-shifted groups, respectively. This is apparently
an indication of maser emission. 
Strikingly, the two main groups of the other
masers behave conversely to those of the
6031\,MHz maser in the relative flux.

The spectrum of the water maser is markedly different from those
of OH masers. We only detect two velocity components at 
116.3 and 121.4\,$\rm km~s^{-1}$, with latter also detected in
the 1612\,MHz maser.
In contrast to the observations of \cite{2007ApJ...664.1130D}, it no longer exhibits an extremely broad 
velocity spread (as shown in the bottom panel of Figure~\ref{fig:oh}). \citet{2023PASJ...75.1183I} performed a recent monitor observation which showed that the high-velocity water maser components faded out
until February 2023.  
Our observations are in line with that.

\begin{figure}[ht!]
\plotone{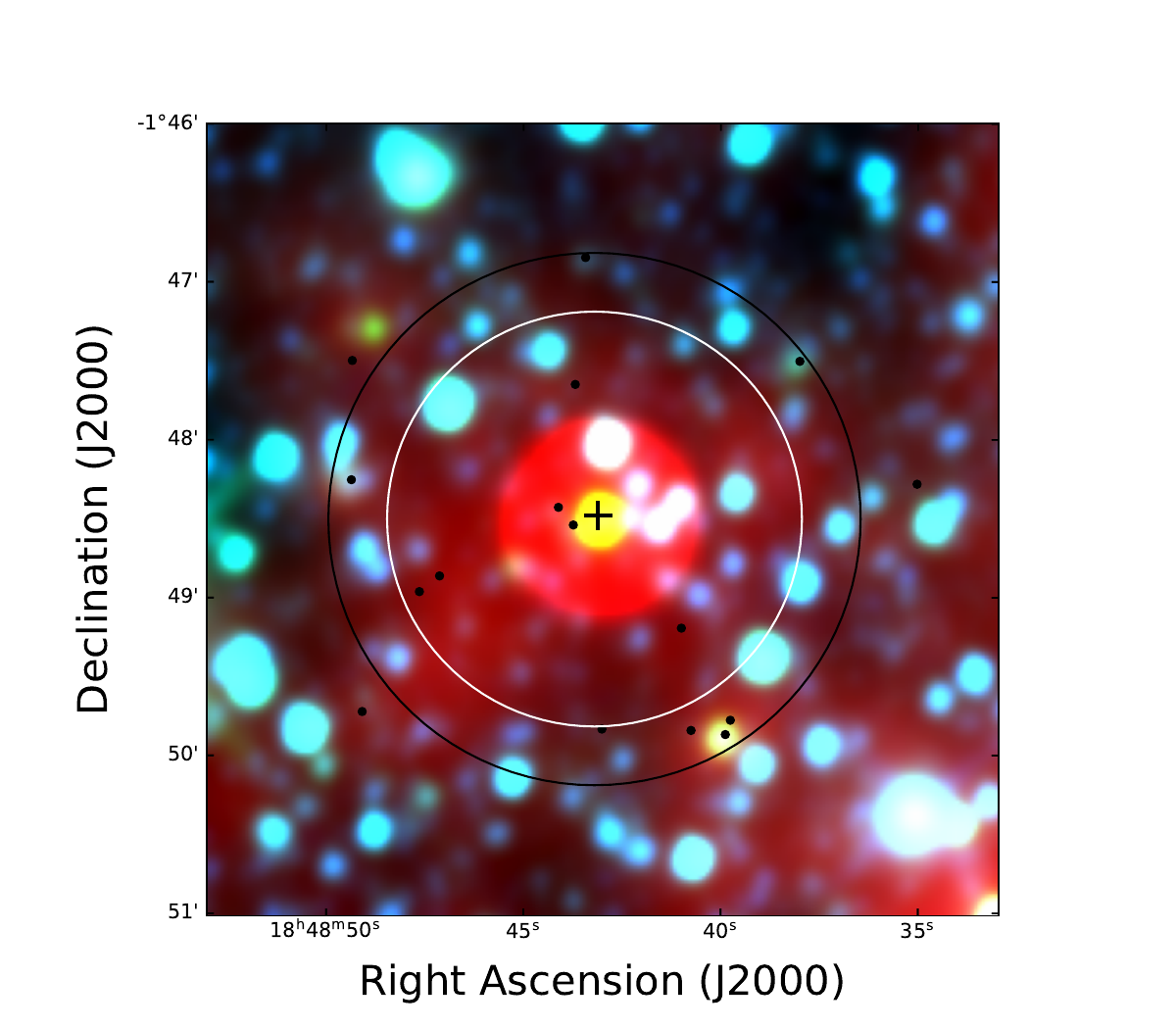}
\caption{The WISE image of I18460. The 3.4, 4.6, and 22\,$\mu m$ bands are 
shown in blue, green, and red, respectively. 
The white and black circles indicate the beam sizes of TMRT at 6.0 and 4.7\,GHz, respectively, and the cross marks the position of the beam center.
The  dots denote the surrounding young stellar objects.
\label{fig:wise}}
\end{figure}

In order to rule out chances from the contamination of nearby sources,
we examined the Wide-field Infrared Survey Explorer (WISE) image.
As shown in  Figure~\ref{fig:wise}, I18460 is the only visible infrared
source within the TMRT beam, and 
 there is no other infrared pumping source. 
The nearest maser source is G031.213-0.180, a HMSFR lying 
$\sim$12$\arcmin$ away from I18460. 
\citet{2022ApJ...928..129Q} discovered the 4660\,MHz ex-OH maser toward this source,
which peaks at 111.1\,$\rm km~s^{-1}$ and has a flux density of 0.3\,Jy.
Therefore, it is impossible to contaminate our detection.

The 4660\,MHz ex-OH  maser has a center velocity  comparable to the systemic velocity obtained from the Very Long Baseline Array \citep[124.7\,$\rm km~s^{-1}$,][]{2013ApJ...773..182I}.
It is likely to originate from an expanding shell with a velocity of
$\sim$14\,$\rm km~s^{-1}$, which is slightly lower than that indicated
by the g-OH 1665\,MHz masers ($\sim$15\,$\rm km~s^{-1}$, see
Figure~\ref{fig:oh}). We hypothesize that the ex-OH 4660\,MHz maser
may traces the  high-density shocked gas which has been decelerated through
the interaction between the stellar winds and the
interstellar medium. A more sophisticated analysis, like that performed
by \cite{2008ApJ...676..371P} for supernova remnants, is required
to elucidate the behavior of the ex-OH masers.
It is intriguing to note that  the 
ex-OH 4660\,MHz and g-OH 1665\,MHz 
masers seem to be complementary in
profile, and all their peaks can 
 be seen in the g-OH 1612\,MHz maser (Figure~\ref{fig:oh}).

Previous observations of  ex-OH masers toward evolved stars
only reveal blue-shifted component
\citep{2010A&A...520A..45D,2019ApJ...878...90S,2020MNRAS.495.4326H}, which 
might be partly
caused by the effect of the optical thickness.
The current work is the  first case of ex-OH masers being discovered exclusively on the red-shifted side.
Previous observations of ex-OH masers 
include the 6035\,MHz in
 Vy\,2-2 and K\,3-35 \citep{2010A&A...520A..45D}
and the 4765\,MHz line in CRL\,618 \citep{2019ApJ...878...90S}. 
Both masers are not discovered in I18460,
suggesting that the excitation of the ex-OH maser is very sensitive to the
circumstellar physical conditions. 
We observed an asymmetry in intensity between the red- and blue-shifted groups in the 4660 MHz  line, as well as in the 6031 MHz  line. This phenomenon could be attributed to 
the spatial fluctuations of density, temperature,
and velocity fields.

Extensive researches have been performed on the pumping of g-OH masers in  CSEs \citep[e.g.][]{1998A&A...331..317T,2005MNRAS.364..783G}, but little is known about the pumping process of ex-OH masers.
The classic model of g-OH maser in circumstellar envelope proposed by \cite{1976ApJ...205..384E} suggests that 
the 1612\,MHz  maser 
can be pumped from the $^2\prod_{1/2}$, $J=5/2$ state by infrared radiation around 35\,$\mu$m.
\citet{1981ASSL...88..363E} suggested that infrared radiation around
53\,$\mu$m could also contribute to the pumping of the 1612\,MHz OH maser
from the $^2\prod_{1/2}$, $J=3/2$ state, which was subsequently found to
be the dominant inversion mechanism  \citep{2005MNRAS.364..783G}.
Apart from that, the collisional excitation within the $^2\prod_{3/2}$ 
ladder also plays an important role on the pumping of this maser 
\citep{2005MNRAS.364..783G}.
Observations of ex-OH masers may shed 
important insight into the model. 
The 4660\,MHz maser is pumped by infrared radiation
at 79\,$\mu$m, and its large intensity indicates
that infrared pumping from $^2\prod_{1/2}$
significantly contributes to the
inversion of the 1612\,MHz g-OH maser level.
The model developed for star-forming regions predicts that 
it is possible to simultaneously 
detect the two masers 
under the conditions of a strong far-infrared line overlap and long gain lengths \citep{1992A&A...262..555G}.

The weakness of the 6031\,MHz maser suggests
that the collisional excitation within the $^2\prod_{3/2}$ ladder may be insignificant
for the excitation of the 1612\,MHz g-OH maser.
I18460 is the only WF detected in ex-OH masers so far.
However, its other properties, such as infrared luminosity and
color, are not peculiar compared to other WFs. 
The origin of intense ex-OH masers in I18460 is unclear.
Further search for ex-OH masers in WFs is desirable.

\begin{figure}[ht!]
\plotone{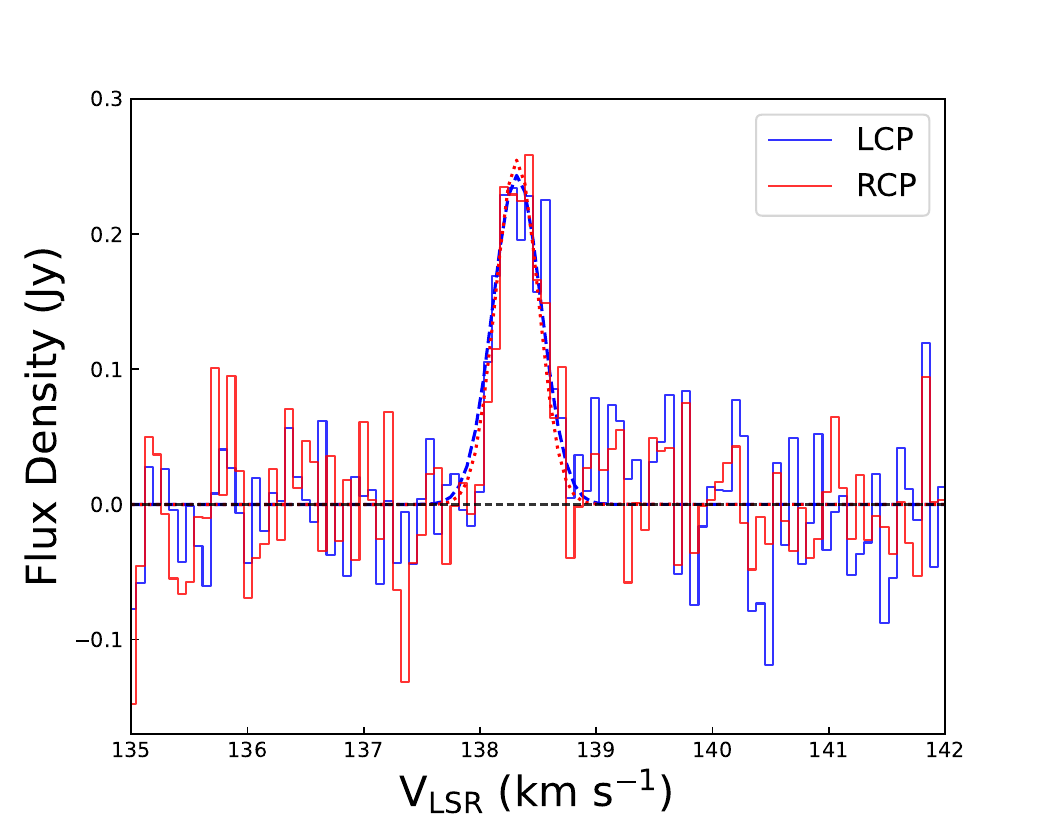}
\caption{ Circular polarization spectra of
the 6031\,MHz OH maser.
The stepped and dashed curves represent the
observations and  Gaussian fittings, respectively.
\label{fig:cp}}
\end{figure}

\citet{2012A&A...537A...5W} did not detect the
Zeeman splitting in the g-OH maser at 1665\,MHz.
We have  attempted to examine the Zeeman splitting by comparing
the LCP and RCP  of the ex-OH masers at 4660\,MHz and 6031\,MHz.
{ Figure~\ref{fig:cp} shows the LCP and RCP
spectra of the 6031\,MHz maser.
They were obtained by co-adding
the spectra taken on different dates, 
resulting in a rms of $\sim50$\,mJy.
Gaussian fittings show that the 
6031\,MHz LCP and RCP lines highly coincide, and both have
 a peak velocity of 138.32\,$\rm km~s^{-1}$ 
and a width of 0.46\,$\rm km~s^{-1}$. 
Therefore, we cannot detect the 
Zeeman splitting under the velocity resolution of 0.07\,$\rm km~s^{-1}$. }
The 4660\,MHz maser has a negligible Lande g-factor
has thus cannot exhibit the Zeeman splitting \citep{2001MNRAS.324..643G}. 
Through the observation of the 6031\,MHz maser, we estimate the upper limit of the magnetic field strength to be $\sim$0.89\,mG. 
Such a weak magnetic field is unlikely to contribute to
the PN shaping. However, it should be cautioned that the OH
shell is far away from the central engine accelerating
the WF jets. Therefore, we cannot completely rule out the existence of a stronger  magnetic field in the central impact region. Further interferometric observations are desirable to 
investigate the role of magnetic fields in PNe shaping.

\section{Conclusion} \label{sec:concl}

We unambiguously detected rare ex-OH masers at 4660 and 6031\,MHz toward the WF source I18460 using the TMRT. To our best knowledge, this is for the
first time that the ex-OH masers are discovered in a WF source and that
the ex-OH maser at 4660\,MHz is observed in CSEs of evolved stars.
The 4660\,MHz maser turns out to be abnormally strong and
closely resembles the g-OH maser at 1612\,MHz. The 6031\,MHz maser 
exhibits the red-shifted component only. The puzzling behaviors of the ex-OH
masers are largely unexplored, and would  call for 
high spatial resolution observations and elaborated
maser modelling.

\section*{Acknowledgements}

{ We acknowledge an anonymous referee for his/her constructive comments.}
The financial supports of this work are from the National Science Foundation of China (NSFC, No.\,12333005 and No. 11973099) and the science research grants from the China Manned Space Project (NO. CMS-CSST-2021-A09, CMS-CSST-2021-A10, etc).
H.-H.Q. is supported by the Youth Innovation Promotion Association CAS and National Natural Science Foundation of China (grant No. 11903038).


\bibliography{sample631}{}

\begin{thebibliography}{}
\expandafter\ifx\csname natexlab\endcsname\relax\def\natexlab#1{#1}\fi
\providecommand{\url}[1]{\href{#1}{#1}}
\providecommand{\dodoi}[1]{doi:~\href{http://doi.org/#1}{\nolinkurl{#1}}}
\providecommand{\doeprint}[1]{\href{http://ascl.net/#1}{\nolinkurl{http://ascl.net/#1}}}
\providecommand{\doarXiv}[1]{\href{https://arxiv.org/abs/#1}{\nolinkurl{https://arxiv.org/abs/#1}}}

\bibitem[{{Amiri} {et~al.}(2012){Amiri}, {Vlemmings}, {Kemball}, \& {van Langevelde}}]{2012A&A...538A.136A}
{Amiri}, N., {Vlemmings}, W.~H.~T., {Kemball}, A.~J., \& {van Langevelde}, H.~J. 2012, \aap, 538, A136, \dodoi{10.1051/0004-6361/201117452}

\bibitem[{{Balick} {et~al.}(2020){Balick}, {Frank}, \& {Liu}}]{2020ApJ...889...13B}
{Balick}, B., {Frank}, A., \& {Liu}, B. 2020, \apj, 889, 13, \dodoi{10.3847/1538-4357/ab5651}

\bibitem[{{Claussen} \& {Fix}(1981)}]{1981ApJ...250L..77C}
{Claussen}, M.~J., \& {Fix}, J.~D. 1981, \apjl, 250, L77, \dodoi{10.1086/183677}

\bibitem[{{Cragg} {et~al.}(2002){Cragg}, {Sobolev}, \& {Godfrey}}]{2002MNRAS.331..521C}
{Cragg}, D.~M., {Sobolev}, A.~M., \& {Godfrey}, P.~D. 2002, \mnras, 331, 521, \dodoi{10.1046/j.1365-8711.2002.05226.x}

\bibitem[{{Deguchi} {et~al.}(2007){Deguchi}, {Nakashima}, {Kwok}, \& {Koning}}]{2007ApJ...664.1130D}
{Deguchi}, S., {Nakashima}, J.-i., {Kwok}, S., \& {Koning}, N. 2007, \apj, 664, 1130, \dodoi{10.1086/519154}

\bibitem[{{Desmurs}(2012)}]{2012IAUS..287..217D}
{Desmurs}, J.~F. 2012, in Cosmic Masers - from OH to H0, ed. R.~S. {Booth}, W.~H.~T. {Vlemmings}, \& E.~M.~L. {Humphreys}, Vol. 287, 217--224, \dodoi{10.1017/S1743921312006990}

\bibitem[{{Desmurs} {et~al.}(2010){Desmurs}, {Baudry}, {Sivagnanam}, {Henkel}, {Richards}, \& {Bains}}]{2010A&A...520A..45D}
{Desmurs}, J.~F., {Baudry}, A., {Sivagnanam}, P., {et~al.} 2010, \aap, 520, A45, \dodoi{10.1051/0004-6361/200913387}

\bibitem[{{Elitzur}(1981)}]{1981ASSL...88..363E}
{Elitzur}, M. 1981, in Astrophysics and Space Science Library, Vol.~88, Physical Processes in Red Giants, ed. J.~{Iben}, I. \& A.~{Renzini}, 363--382, \dodoi{10.1007/978-94-009-8492-9_40}

\bibitem[{{Elitzur} {et~al.}(1976){Elitzur}, {Goldreich}, \& {Scoville}}]{1976ApJ...205..384E}
{Elitzur}, M., {Goldreich}, P., \& {Scoville}, N. 1976, \apj, 205, 384, \dodoi{10.1086/154289}

\bibitem[{{Falceta-Gon{\c{c}}alves} \& {Jatenco-Pereira}(2002)}]{2002ApJ...576..976F}
{Falceta-Gon{\c{c}}alves}, D., \& {Jatenco-Pereira}, V. 2002, \apj, 576, 976, \dodoi{10.1086/341794}

\bibitem[{{Frank} \& {Blackman}(2004)}]{2004ApJ...614..737F}
{Frank}, A., \& {Blackman}, E.~G. 2004, \apj, 614, 737, \dodoi{10.1086/382018}

\bibitem[{{Gardner} {et~al.}(1987){Gardner}, {Whiteoak}, \& {Palmer}}]{1987MNRAS.225..469G}
{Gardner}, F.~F., {Whiteoak}, J.~B., \& {Palmer}, P. 1987, \mnras, 225, 469, \dodoi{10.1093/mnras/225.3.469}

\bibitem[{{Gledhill} {et~al.}(2001){Gledhill}, {Chrysostomou}, {Hough}, \& {Yates}}]{2001MNRAS.322..321G}
{Gledhill}, T.~M., {Chrysostomou}, A., {Hough}, J.~H., \& {Yates}, J.~A. 2001, \mnras, 322, 321, \dodoi{10.1046/j.1365-8711.2001.04112.x}

\bibitem[{{G{\'o}mez} {et~al.}(2011){G{\'o}mez}, {Rizzo}, {Su{\'a}rez}, {Miranda}, {Guerrero}, \& {Ramos-Larios}}]{2011ApJ...739L..14G}
{G{\'o}mez}, J.~F., {Rizzo}, J.~R., {Su{\'a}rez}, O., {et~al.} 2011, \apjl, 739, L14, \dodoi{10.1088/2041-8205/739/1/L14}

\bibitem[{{G{\'o}mez} {et~al.}(2017){G{\'o}mez}, {Su{\'a}rez}, {Rizzo}, {Uscanga}, {Walsh}, {Miranda}, \& {Bendjoya}}]{2017MNRAS.468.2081G}
{G{\'o}mez}, J.~F., {Su{\'a}rez}, O., {Rizzo}, J.~R., {et~al.} 2017, \mnras, 468, 2081, \dodoi{10.1093/mnras/stx650}

\bibitem[{{G{\'o}mez} {et~al.}(2016){G{\'o}mez}, {Uscanga}, {Green}, {Miranda}, {Su{\'a}rez}, \& {Bendjoya}}]{2016MNRAS.461.3259G}
{G{\'o}mez}, J.~F., {Uscanga}, L., {Green}, J.~A., {et~al.} 2016, \mnras, 461, 3259, \dodoi{10.1093/mnras/stw1536}

\bibitem[{{Gray} {et~al.}(2001){Gray}, {Cohen}, {Richards}, {Yates}, \& {Field}}]{2001MNRAS.324..643G}
{Gray}, M.~D., {Cohen}, R.~J., {Richards}, A.~M.~S., {Yates}, J.~A., \& {Field}, D. 2001, \mnras, 324, 643, \dodoi{10.1046/j.1365-8711.2001.04349.x}

\bibitem[{{Gray} {et~al.}(1992){Gray}, {Field}, \& {Doel}}]{1992A&A...262..555G}
{Gray}, M.~D., {Field}, D., \& {Doel}, R.~C. 1992, \aap, 262, 555

\bibitem[{{Gray} {et~al.}(2005){Gray}, {Howe}, \& {Lewis}}]{2005MNRAS.364..783G}
{Gray}, M.~D., {Howe}, D.~A., \& {Lewis}, B.~M. 2005, \mnras, 364, 783, \dodoi{10.1111/j.1365-2966.2005.09591.x}

\bibitem[{{Habing}(1996)}]{1996A&ARv...7...97H}
{Habing}, H.~J. 1996, \aapr, 7, 97, \dodoi{10.1007/PL00013287}

\bibitem[{{Herman} {et~al.}(1985){Herman}, {Baud}, \& {Habing}}]{1985A&A...144..514H}
{Herman}, J., {Baud}, B., \& {Habing}, H.~J. 1985, \aap, 144, 514

\bibitem[{{Hou} \& {Gao}(2020)}]{2020MNRAS.495.4326H}
{Hou}, L.~G., \& {Gao}, X.~Y. 2020, \mnras, 495, 4326, \dodoi{10.1093/mnras/staa1461}

\bibitem[{{Imai}(2007)}]{2007IAUS..242..279I}
{Imai}, H. 2007, in Astrophysical Masers and their Environments, ed. J.~M. {Chapman} \& W.~A. {Baan}, Vol. 242, 279--286, \dodoi{10.1017/S1743921307013130}

\bibitem[{{Imai} {et~al.}(2013){Imai}, {Deguchi}, {Nakashima}, {Kwok}, \& {Diamond}}]{2013ApJ...773..182I}
{Imai}, H., {Deguchi}, S., {Nakashima}, J.-i., {Kwok}, S., \& {Diamond}, P.~J. 2013, \apj, 773, 182, \dodoi{10.1088/0004-637X/773/2/182}

\bibitem[{{Imai} {et~al.}(2020){Imai}, {Uno}, {Maeyama}, {Yamaguchi}, {Amada}, {Hamae}, {Orosz}, {G{\'o}mez}, {Tafoya}, {Uscanga}, \& {Burns}}]{2020PASJ...72...58I}
{Imai}, H., {Uno}, Y., {Maeyama}, D., {et~al.} 2020, \pasj, 72, 58, \dodoi{10.1093/pasj/psaa047}

\bibitem[{{Imai} {et~al.}(2023){Imai}, {Hamae}, {Amada}, {Nakashima}, {Shum}, {Kasai}, {G{\'o}mez}, {Uscanga}, {Tafoya}, {Orosz}, \& {Burns}}]{2023PASJ...75.1183I}
{Imai}, H., {Hamae}, Y., {Amada}, K., {et~al.} 2023, \pasj, 75, 1183, \dodoi{10.1093/pasj/psad064}

\bibitem[{{Khouri} {et~al.}(2021){Khouri}, {Vlemmings}, {Tafoya}, {P{\'e}rez-S{\'a}nchez}, {S{\'a}nchez Contreras}, {G{\'o}mez}, {Imai}, \& {Sahai}}]{2022NatAs...6..275K}
{Khouri}, T., {Vlemmings}, W. H.~T., {Tafoya}, D., {et~al.} 2021, Nature Astronomy, 6, 275, \dodoi{10.1038/s41550-021-01528-4}

\bibitem[{{Kwok}(1993)}]{1993ARA&A..31...63K}
{Kwok}, S. 1993, \araa, 31, 63, \dodoi{10.1146/annurev.aa.31.090193.000431}

\bibitem[{{Li} {et~al.}(2016){Li}, {Shen}, {Wang}, {Chen}, {Wu}, {Zhao}, {Wang}, {Zuo}, {Fan}, {Hong}, {Jiang}, {Li}, {Liang}, {Ling}, {Liu}, {Qian}, {Zhang}, {Zhong}, \& {Ye}}]{2016ApJ...824..136L}
{Li}, J., {Shen}, Z.-Q., {Wang}, J., {et~al.} 2016, \apj, 824, 136, \dodoi{10.3847/0004-637X/824/2/136}

\bibitem[{{Livio} \& {Soker}(1988)}]{1988ApJ...329..764L}
{Livio}, M., \& {Soker}, N. 1988, \apj, 329, 764, \dodoi{10.1086/166419}

\bibitem[{{Nordhaus} \& {Blackman}(2006)}]{2006MNRAS.370.2004N}
{Nordhaus}, J., \& {Blackman}, E.~G. 2006, \mnras, 370, 2004, \dodoi{10.1111/j.1365-2966.2006.10625.x}

\bibitem[{{Pety}(2005)}]{2005sf2a.conf..721P}
{Pety}, J. 2005, in SF2A-2005: Semaine de l'Astrophysique Francaise, ed. F.~{Casoli}, T.~{Contini}, J.~M. {Hameury}, \& L.~{Pagani}, 721

\bibitem[{{Pihlstr{\"o}m} {et~al.}(2008){Pihlstr{\"o}m}, {Fish}, {Sjouwerman}, {Zschaechner}, {Lockett}, \& {Elitzur}}]{2008ApJ...676..371P}
{Pihlstr{\"o}m}, Y.~M., {Fish}, V.~L., {Sjouwerman}, L.~O., {et~al.} 2008, \apj, 676, 371, \dodoi{10.1086/529009}

\bibitem[{{Qiao} {et~al.}(2022){Qiao}, {Shen}, {Breen}, {Yang}, {Chen}, \& {Li}}]{2022ApJ...928..129Q}
{Qiao}, H.-H., {Shen}, Z.-Q., {Breen}, S.~L., {et~al.} 2022, \apj, 928, 129, \dodoi{10.3847/1538-4357/ac5820}

\bibitem[{{Rizzo} {et~al.}(2013){Rizzo}, {G{\'o}mez}, {Miranda}, {Osorio}, {Su{\'a}rez}, \& {Dur{\'a}n-Rojas}}]{2013A&A...560A..82R}
{Rizzo}, J.~R., {G{\'o}mez}, J.~F., {Miranda}, L.~F., {et~al.} 2013, \aap, 560, A82, \dodoi{10.1051/0004-6361/201322187}

\bibitem[{{Sahai} \& {Trauger}(1998)}]{1998AJ....116.1357S}
{Sahai}, R., \& {Trauger}, J.~T. 1998, \aj, 116, 1357, \dodoi{10.1086/300504}

\bibitem[{{Sevenster} {et~al.}(1997){Sevenster}, {Chapman}, {Habing}, {Killeen}, \& {Lindqvist}}]{1997A&AS..124..509S}
{Sevenster}, M.~N., {Chapman}, J.~M., {Habing}, H.~J., {Killeen}, N.~E.~B., \& {Lindqvist}, M. 1997, \aaps, 124, 509, \dodoi{10.1051/aas:1997365}

\bibitem[{{Sjouwerman} {et~al.}(2007){Sjouwerman}, {Fish}, {Claussen}, {Pihlstr{\"o}m}, \& {Zschaechner}}]{2007ApJ...666L.101S}
{Sjouwerman}, L.~O., {Fish}, V.~L., {Claussen}, M.~J., {Pihlstr{\"o}m}, Y.~M., \& {Zschaechner}, L.~K. 2007, \apjl, 666, L101, \dodoi{10.1086/521827}

\bibitem[{{Strack} {et~al.}(2019){Strack}, {Araya}, {Lebr{\'o}n}, {Minchin}, {Arce}, {Ghosh}, {Hofner}, {Kurtz}, {Olmi}, {Pihlstr{\"o}m}, \& {Salter}}]{2019ApJ...878...90S}
{Strack}, A., {Araya}, E.~D., {Lebr{\'o}n}, M.~E., {et~al.} 2019, \apj, 878, 90, \dodoi{10.3847/1538-4357/ab1f93}

\bibitem[{{Su{\'a}rez} {et~al.}(2008){Su{\'a}rez}, {G{\'o}mez}, \& {Miranda}}]{2008ApJ...689..430S}
{Su{\'a}rez}, O., {G{\'o}mez}, J.~F., \& {Miranda}, L.~F. 2008, \apj, 689, 430, \dodoi{10.1086/592493}

\bibitem[{{Tafoya} {et~al.}(2020){Tafoya}, {Imai}, {G{\'o}mez}, {Nakashima}, {Orosz}, \& {Yung}}]{2020ApJ...890L..14T}
{Tafoya}, D., {Imai}, H., {G{\'o}mez}, J.~F., {et~al.} 2020, \apjl, 890, L14, \dodoi{10.3847/2041-8213/ab70b8}

\bibitem[{{Thai-Q-Tung} {et~al.}(1998){Thai-Q-Tung}, {Dinh-v-Trung}, {Nguyen-Q-Rieu}, {Bujarrabal}, {Le Bertre}, \& {Gerard}}]{1998A&A...331..317T}
{Thai-Q-Tung}, {Dinh-v-Trung}, {Nguyen-Q-Rieu}, {et~al.} 1998, \aap, 331, 317

\bibitem[{{Tocknell} {et~al.}(2014){Tocknell}, {De Marco}, \& {Wardle}}]{2014MNRAS.439.2014T}
{Tocknell}, J., {De Marco}, O., \& {Wardle}, M. 2014, \mnras, 439, 2014, \dodoi{10.1093/mnras/stu079}

\bibitem[{{Ueta} {et~al.}(2000){Ueta}, {Meixner}, \& {Bobrowsky}}]{2000ApJ...528..861U}
{Ueta}, T., {Meixner}, M., \& {Bobrowsky}, M. 2000, \apj, 528, 861, \dodoi{10.1086/308208}

\bibitem[{{Uscanga} {et~al.}(2023){Uscanga}, {Imai}, {G{\'o}mez}, {Tafoya}, {Orosz}, {McCarthy}, {Hamae}, \& {Amada}}]{2023ApJ...948...17U}
{Uscanga}, L., {Imai}, H., {G{\'o}mez}, J.~F., {et~al.} 2023, \apj, 948, 17, \dodoi{10.3847/1538-4357/acc06f}

\bibitem[{{Vlemmings}(2014)}]{2014IAUS..302..389V}
{Vlemmings}, W.~H.~T. 2014, in Magnetic Fields throughout Stellar Evolution, ed. P.~{Petit}, M.~{Jardine}, \& H.~C. {Spruit}, Vol. 302, 389--397, \dodoi{10.1017/S1743921314002580}

\bibitem[{{Wardle}(2007)}]{2007IAUS..242..336W}
{Wardle}, M. 2007, in Astrophysical Masers and their Environments, ed. J.~M. {Chapman} \& W.~A. {Baan}, Vol. 242, 336--337, \dodoi{10.1017/S1743921307013300}

\bibitem[{{Wolak} {et~al.}(2012){Wolak}, {Szymczak}, \& {G{\'e}rard}}]{2012A&A...537A...5W}
{Wolak}, P., {Szymczak}, M., \& {G{\'e}rard}, E. 2012, \aap, 537, A5, \dodoi{10.1051/0004-6361/201117263}

\bibitem[{{Zuckerman} {et~al.}(1972){Zuckerman}, {Yen}, {Gottlieb}, \& {Palmer}}]{1972ApJ...177...59Z}
{Zuckerman}, B., {Yen}, J.~L., {Gottlieb}, C.~A., \& {Palmer}, P. 1972, \apj, 177, 59, \dodoi{10.1086/151686}

\end{thebibliography}
\bibliographystyle{aasjournal}



\newpage

\begin{deluxetable*}{ccccccc}[htb]
\tabletypesize{\scriptsize}
\tablewidth{0pt}
\linespread{0.8}
\tablecaption{Measurements of masers in I18460. \label{tab:pro}}
\tablehead{\\
\colhead{Frequency} & \colhead{$\int S_v dv$} & \colhead{V$_{\rm p}$} &  \colhead{Range} & \colhead{S$_{\rm p}$} & \colhead{rms} & \colhead{T$_{\rm b}$}\\
\colhead{(MHz)} & \colhead{($\rm Jy~km~s^{-1}$)} & \colhead{($\rm km~s^{-1}$)} & \colhead{($\rm km~s^{-1}$)} & \colhead{(Jy)} & \colhead{(Jy)} & \colhead{($10^5$~K)}\\
\colhead{(1)} & \colhead{(2)} & \colhead{(3)} & \colhead{(4)} & \colhead{(5)} & \colhead{(6)} & \colhead{(7)}
}
\startdata
        1612 & $171.14$  & 111.1 & $110.0,114.5$ & 123.42 & 0.39 & $>629.77$ \\
        1612 & $10.36$  & 121.1 & $119.0,124.0$ & 6.82 & 0.39 & $>34.80$ \\
        1612 & $59.74$  & 138.4 & $135.8,142.8$ & 28.20 & 0.39 & $>143.90$ \\
        1665 & $8.86$  & 110.4 & $109.0,111.3$ & 20.36 & 0.50 & $>97.38$ \\
        1665 & $4.98$  & 140.4 & $139.2,142.6$ & 4.22 & 0.50 & $>20.18$ \\  
        4660 & $117.86$  & 111.4 & $109.8,117.7$ & 103.40 & 0.16 & $>63.14$ \\  
        4660 & $22.98$  & 138.3 & $135.8,140.7$ & 19.96 & 0.16 & $>12.19$ \\ 
        6031 & $0.18$  & 138.4 & $137.8,138.9$ & 0.52 & 0.09 & $>0.19$ \\ 
        22235 & $1.58$  & 116.3 & $114.9,117.3$ & 2.54 & 0.39 & $>0.07$ \\
        22235 & $3.98$  & 121.4 & $120.2,123.3$ & 4.70 & 0.39 & $>0.13$ \\
\enddata
\end{deluxetable*}

\end{document}